\documentclass{kapproc}
\usepackage{procps}
\usepackage[latin1]{inputenc}
\usepackage{setspace}
\usepackage{epsfig}
\usepackage[english]{babel}
\usepackage{xspace}
\usepackage{url}
\usepackage{amssymb}

\let\footnote\savefootnote
\let\footnotetext\savefootnotetext 
\let\footnoterule\savefootnoterule 

\def\rcs$#1${#1}
\newcommand{\version}[1]{\thanks{\rcs#1}}
\newcommand{\eg}{e.g.,\xspace}
\newcommand{\ie}{i.e.,\xspace}

\newcommand{\aka}[1]{aka.\xspace\textit{#1}}
\newcommand{\sass}{{S \rightarrow S}}
\newcommand{\sash}{{S \rightarrow H}}
\newcommand{\hass}{{H \rightarrow S}}
\newcommand{\hash}{{H \rightarrow H}}

\def\rrefstrip#1.#2\relax{#2}
\newcommand{\rref}[1]{\ref{#1}}

\normallatexbib

\begin{document}

\articletitle[Spam filter analysis]{Spam filter analysis\version{$Id: spam-filter.tex,v 1.24 2004/02/05 12:01:52 jhh Exp $}}

\author{Flavio D. Garcia}
\affil{Department of Computer Science, University of Nijmegen,
  the Netherlands}
\email{flaviog@cs.kun.nl}

\author{Jaap-Henk Hoepman}
\affil{Department of Computer Science, University of Nijmegen,
  the Netherlands}
\email{jhh@cs.kun.nl}

\author{Jeroen van Nieuwenhuizen}
\affil{Department of Computer Science, University of Twente,
  the Netherlands}

\bibliographystyle{plain}

\begin{abstract}
Unsolicited bulk email (\aka{spam}) is a major problem on the
Internet. To counter spam, several techniques, ranging from spam
filters to mail protocol extensions like \emph{hashcash}, have
been proposed. In this paper we investigate the effectiveness of
several spam filtering techniques and technologies. Our analysis
was performed by simulating email traffic under different
conditions. We show that genetic algorithm based spam filters
perform best at server level and na\"{i}ve Bayesian filters are the
most appropriate for filtering at user level.
\end{abstract}

\begin{keywords}
Spam, unsolicited email, email abuse, security, networking, Bayesian
filtering, genetic algorithms, text classification.
\end{keywords}

\section{Introduction}

Spam~\cite{Python89}, officially called unsolicited bulk email (UBE) or
unsolicited commercial email (UCE), is rapidly becoming a major problem on the
Internet. At
the end of 2002, as much as 40\% of all email traffic consisted of spam%
\footnote{\url{http://zdnet.com.com/2100-1106-955842.html}}${}^,$%
\footnote{\url{http://www.linuxsecurity.com/articles/privacy_article-6369.html}},
 and recent reports estimate that this amount has risen to more than
50\%\footnote{\url{http://zdnet.com.com/2100-1105_2-1019528.html}}.

To handle this increasing load of junk email~\cite{postel1975junkmail},
several spam filtering techniques exist to automatically classify incoming
email as spam, and to reject or discard email classified as
such~\cite{StopSpam}. 
In this paper we investigate the effectiveness of these spam
filtering techniques and technologies. 

For users, receiving spam is quite a nuisance and costs money. In
a recent study of the European
Community~\cite{unsolicitedccanddp}, it was estimated that the
cost for receiving spam for an average Internet user is in the
order of 30 euro a year. But the costs of spam goes well beyond
the total costs of all recipients. Each ISP pays for each email
message received, because it must be stored in a mail box and it
takes up a certain amount of bandwidth. The total cost has been
estimated in the order of 10 billion euro a
year~\cite{unsolicitedccanddp}.

A second problem with spam is the impact it has on the Internet
backbone. Spam sent over the Internet backbone causes delays for
all Internet users. Furthermore, because most spammers use
mailing lists that have outdated addresses on them, many messages
are rejected (``bounced''). This mandates the operator of the
intended destination to send a return response, wasting even more
bandwidth~\cite{dontspew}.

Bulk mailers use several different techniques to send their spam.
Often, bulk mailers misuse the SMTP protocol or use badly configured
MTAs (so-called open-relays) to hide their
tracks
\cite{unsolicitedccanddp} \cite{rfc2505}. We describe
these techniques in detail in section~\rref{ssec-bulk}.

There are at least three fundamentally different ways to counter
spammers \cite{spam}.
First, bulk mailers can be
prevented to send spam by blocking or limiting access to mail
servers. Another method is make spamming less profitable, for
example by incurring a cost on every email message
sent~\cite{naor95pricing}. A third method aims to detect and
remove all spam once it is sent by applying different types of
filtering techniques that use the special characteristics of spam
to recognise
it~\cite{EvaluationNaiveBayesian} 
\cite{androutsopoulos-learning} \cite{gabber98curbing}
\cite{channelhall} \cite{sahami98bayesian}.
These techniques are discussed in section~\rref{ssec-counter}.

Our analysis of countermeasures against spam focuses on filtering
techniques. We are interested in measuring the accuracy level of
these filters in practice. Some of the filtering techniques not
only look at the content of each message, but also consider the
email traffic at large (\eg methods that try to detect duplicate
mail messages, or checksum schemes that match incoming messages
with a database of known spam messages). To faithfully analyse
such spam filters, we built a simulator to generate realistic
email traffic and test the filters with it. In
section~\rref{sec-analysis} we give a description of our analysis
method. We have analysed the performance of the filters in two
settings: while being used at the server level and while being
used by the end user directly. While running at server level, the
filter might use the information about connections from the
server. On the other hand, while running at user level, the filter
is able to be trained or customised as a function of user specific
characteristics. Moreover, we have measured the different behaviour
of filters depending on the type of bulk mailer used to generate
the spam. We refer to section~\rref{sec-analysis} for details.

To summarise our results: we have found that filters based on genetic
algorithms, perform best at ISP level and na\"{\i}ve Bayesian
filters perform best at user level. We discuss the results of our
analysis on a per filter basis in section~\rref{sec-results}.

\section{Spam: producers and countermeasures}
In this section we describe the most common techniques used by
bulk mailers. We also describe current proposals for countering
spam, with focus on filtering techniques.

\subsection{Bulk mailing techniques}
\label{ssec-bulk}%
Spammers use so-called bulk mailers to send spam. These bulk
mailers are capable of sending huge volumes of email without
going through a specific mail server or a particular ISP. Some
bulk mailers are capable of sending approximately 250,000 messages
an hour over a 28.8kb$/$s modem line~\cite{unsolicitedccanddp}.
This enormous amount of messages is attained by contacting more
than one mail server at the same time and misusing the resources
of the ISPs. 

The bulk mailers used by spammers have several features to hide
their tracks.  Most bulk mailers do not use the mail server of
their ISP, but instead connect to the destination mail
server directly or use a so-called open relay.  This way, the
spammer avoids to be detected by his ISP. An open relay is a SMTP
or ESMTP server that allows everyone to use that server to relay
mail. To make the tracking even harder when an open relay is used,
most bulk mailers add so-called bogus received headers to the spam
message (in front of the \emph{real} received headers added by the
SMTP protocol).  By adding these bogus headers they hope to 
redirect any tracking
to a site in the fake header. 

Bulk mailers also include features which try to outsmart spam
filters. The most processor consuming feature is to personalise
every message for a recipient. This personalisation of messages
can be classified into two types. In the first type, the spammer
only uses the victim's mail address as the recipients address
instead of using the \verb"Bcc:" headers to send the message to
recipients. In the second type, he also uses the name or mail
address of the victim to personalise the body of the message. Less
processor consuming techniques include the randomisation of the
\verb"Subject:" field and the \verb"From:" address line. Some bulk
mailers also forge the Message-ID and/or do not send the
\verb"To:" header in the SMTP session. Another technique especially
developed to confuse Bayesian filters, is to add extraneous text
in the body of a message. This text is usually a set of
randomly selected words from a dictionary, or just some paragraphs
from news or books.


\subsection{Countermeasures}
\label{ssec-counter}
There are two fundamentally different methods to counter spam. The
first method tries to prevent bulk mailers to send spam, \eg by
incurring a cost on every email message sent, or by blocking or
limiting access to mail servers for spammers. The second method
aims to detect and remove all spam once it is sent, by applying
different types of filtering techniques.

\subsubsection{Spam prevention}
The most direct way to prevent spam is to close all open relays on
the Internet, and to strengthen the SMTP protocol to disallow
bogus received headers and require sender authentication, to
facilitate bulk mailer tracking.  This forces bulk mailers to send
spam through their own ISP, but relies on these ISPs to block
their accounts. More fundamentally, this approach goes against the
open philosophy of the Internet and poses yet another threat to
privacy on the Internet. Moreover closing all open relays would
not be sufficient, as spammers are now shifting to the use of open
proxies to hide their
tracks\footnote{\url{http://zdnet.com.com/2100-1106-958847.html}},
or even use hacked computers.

One interesting and perhaps more feasible approach proposed to
stop junk email is by using economic based solutions. The
principal attractiveness of spam is that sending large amounts of
small email messages is relatively cheap compared to other direct
marketing techniques. The idea behind such economic based methods
is to make the sending of email more expensive, thereby making it
less attractive to send huge amounts of mail. The two main
categories of economic solutions are computing time based systems
(that force the spammer to spend considerable amounts of his
computing resources to send a single spam message) and money based
systems (that charge a small amount of money for every email
sent).

\paragraph{Computing time based systems}
In computing time based economic solution, the sender of a message
is required to compute a moderately expensive function. This
function is called a \emph{pricing
function}~\cite{hird-technical} \cite{back-hashcash}.
The idea is that for a legitimate sender it is not too expensive
in computer time to send a message to a recipient, but for a
spammer, who has to send many messages, it is. This
expensive use of computer time makes it much harder for a
spammer to send large volumes of mail within an
acceptable time.

The question whether such a system will work in practice is hard
to answer. First of all, this feature has to be incorporated into
the Internet, and this may not be easy (although, admittedly, it
does not require changes to the SMTP protocol and could simply be
enforced by mail user agents). Second, there is the problem of
hardware backward compatibility. A user using an old computer must
be able to send an email in a reasonable amount of time. This
rules out the use of too costly pricing functions. But then for a
spammer using modern hardware, the cost in time to send a message
may become almost equal to zero. It seems impossible to find a
pricing function that suits both needs (although methods based on memory
bandwidth limitations appear a little more promising~\cite{abadi-moderately}).

\paragraph{Money based systems}
Money based systems are based on channelised email systems where
users require payment before reading message arriving on certain
channels~\cite{hird-technical}. This payment can be in the form of
electronic cash to automate the process. This approach would make
it more costly to send Junk-E-mail, which makes it less
attractive.

Problems with money based systems include (among others) the
adoption of the system by users, and the absence of a global
electronic cash system. It is therefore hard, if not impossible,
in practice to introduce a money based spam prevention system.

\subsubsection{Spam Filters}
Filter based countermeasures against spam can be divided into two
main categories: cooperative filtering and heuristic filtering.
The main principle of the former is that there can be
cooperation between spam originators and spam recipients. Such a
cooperative filtering system requires a network-wide
implementation of, and adherence to, a set of standards for
identifying spam. Because most spammers try to hide to track such
an implementation is not likely to appear, hence we will focus on
heuristic filtering.

Heuristic filters work on the assumption that it is possible to
distinguish between normal mail and spam by applying heuristic
rules to a message. We can distinguish three types of heuristic
rules: origin based filtering, filtering based on traffic analysis
and content-based filtering.

\paragraph{Origin based filtering}
Origin based filtering happens before a message is fully received
by the computer of the recipient. The most prominent method in
this class uses the so-called blacklists (\eg the blacklist from
ordb.org\footnote{\url{http://www.ordb.org/}}). These blacklist
can be used to refuse IP or TCP connections from spam originators,
but also to refuse mail if the domain name given at the
\verb"FROM:" command is on the blacklist. 
This method
can be circumvented by relaying mail through the
SMTP servers of legitimate originators that are not on the
blacklist. Another disadvantage is that this sites are frequently
under denial of service attacks (\eg the blacklist of
osirusoft.com). 

A second approach is to configure the SMTP server to perform a
reverse DNS lookup to find the IP-address associated with the
domain name given at the \verb"MAIL FROM" command. If this
IP-address is not equal to the IP-address of the TCP-connection,
the SMTP server can refuse to handle the message. This method can
also be circumvented by using relay hosts.

Whitelists are another origin based filter method. 
They contain the senders (or domains) from which
incoming mail is automatically accepted for delivery. All other
mail is refused by default. To enable legitimate senders to reach
the recipient, a whitelist based system will return a request for
confirmation to the sender, who should reply to this message
within a short period of time. When a whitelist is used it is
almost certain that no spam will reach the inbox of the user. The
disadvantage, however, is that a whitelist has a very high false
positive rate, which may seriously confuse or irritate unknown but
legitimate senders. Moreover, bulk mailers increasingly match the
from header in their mails to the domain of the recipient, using
known (and probably trusted) senders in the same domain.

\paragraph{Filtering based on traffic analysis}
Traffic analysis based filtering can be used at the mail server of
the ISP. Here the log files of the SMTP server can be used to
detect anomalies in the normal traffic stream. Anomalies that can
appear and indicate the spam are anomalies in the connection time
to the server, anomalies in the amount of mail coming from a
certain host, the fact that a message is send to more recipients
as normal from a certain host or the fact that mail is relayed.

\paragraph{Content-based filtering}
Content-based filtering happens after a message is fully received
(including the body of the message). In this case, filtering can
also be based on known keywords in the subject and body of the
message, common features of spam and the use of
signature/checksums from databases on the internet.

\newcommand{\Spam}{\mathbf{S}}
\newcommand{\Ham}{\mathbf{H}}

\emph{Na\"{\i}ve Bayesian filtering} is a new content-based
mechanism for spam
filtering~\cite{sahami98bayesian} \cite{EvaluationNaiveBayesian} \cite{graham}.
Before it can be used, a Bayesian filter must be trained with a
set of spam and a set of legitimate emails (\aka{ham}) that have
been previously classified. For each word $w$ in the training sets
the filter estimates the probability that it occurs in a spam message
($C=\Spam$) or in a ham message ($C=\Ham$) using 
\[
P(W=w|C=\Spam)=\frac{S(w)/N_S}{S(w)/N_S+H(w)/N_H},
\]
where
$S(x)$ is the number of occurrences of word $x$ in the spam set,
 $H(x)$ is the number of occurrences of word $x$ in the ham set, and
 $N_S$ and $N_H$ are the sizes of the spam and ham training
  sets respectively.

When a new message $M=\{w_1,...,w_N\}$ arrives, the filter
determines the $n$ most \emph{interesting} words 
$\{\dot{w}_1 \ldots \dot{w}_n\} \subseteq M$, where interesting means 
$P(\dot{w}_i)\approx 1$ or $P(\dot{w}_i)\approx 0$). 
Using these `interesting' words, the filter then
computes the probability that $M$ is spam using Bayes rule~\cite{Feller71}
\cite{androutsopoulos-learning}
\[
P(C=\Spam|\overrightarrow{W} = M) = 
 \frac{P(C = \Spam) \prod_{i} P(W =\dot{w}_i|C = \Spam)}
      {\sum_{k \in \{\Spam,\Ham\}} P(C = k) 
               \prod_{i} P(W =\dot{w}_i|C = k)}
\]
Then if this probability is greater than a given threshold $T$,
$M$ is classified as spam.

A third method to filter on content is the use of \emph{genetic
algorithms}\footnote{\url{http://spamassassin.org/}}. These
genetic algorithms use so-called feature detectors to score an
email message. This score can then be used to classify a message
as spam or ham.

Another approach which has the ability to learn is the use of
\emph{neural networks}. Like Bayesian filters, neural networks
must be trained first on a set of spam and non-spam messages.
After this training the neural network can be used to classify
incoming mail message based on common features in email
messages\cite{rich02neural}.

It is also possible to classify mail based on the content by the
use of \emph{signature/checksum
schemes}\footnote{\url{http://www.rhyolite.com/anti-spam/dcc/}}
in a cooperative system. When a mail message arrives a
signature/checksum is calculated for this message and compared to
the values in special spam databases on the Internet. If the
checksum matches any of the values in the database, the message is regarded
as spam.

\section{Method of analysis}
\label{sec-analysis}%
In this section we detail the method of our analysis. Our goal is
to measure the effectiveness of several spam-filtering techniques
on realistic email traffic patterns, both at ISP and at the user
level. This traffic should reflect the fact that different users
have different traffic patterns and their emails also differ in
their contents. In an ideal situation we would perform these
measurements on real email traffic, but due to privacy reasons,
that is infeasible (unless we restrict attention to a very small
and atypical consenting sample like staff at a university). 
Instead we perform this analysis by generating both normal and spam
email traffic using a simulator, and measuring the accuracy level of each
analysed spam filter.

\subsection{Mechanism of the analysis}
Our analysis aims to measure how good a spam filter is at
preventing spam from being delivered to the end user, while still
allowing legitimate emails to pass through unblocked. To this
end, we compute for each spam filter, the false acceptance and
rejection rate. Moreover, we propose a single measure of the
filter's wrongness as a function of its false acceptance and
rejection rates. Using this measure, we can rank the accuracy of
the evaluated filters. We also study how the spam personalisation
and new techniques like the inclusion of random words affects this
performance.

For each message, a spam filter does one of four things:
\begin{description}
\item[$\sass$:] it correctly classifies a spam message as spam,
\item[$\sash$:] it falsely accepts a spam message as ham,
\item[$\hash$:] it correctly classifies a ham message as
legitimate, or
\item[$\hass$:] it falsely rejects a ham messages
as spam.
\end{description}
When running the spam filter on $n$ messages, $n_S$ of which are
spam messages and the remaining $n_H$ are legitimate messages, we
write $n_\sass$ for the number of correctly classified spam
messages, and define $n_\sash$, $n_\hass$ and $n_\hash$
analogously. Then the false acceptance rate (FAR) and the false
rejection rate (FRR) are defined as
\[
FAR = \frac{n_\sash}{n_S} \qquad FRR = \frac{n_\hass}{n_H}~.
\]

Clearly, the false acceptance rate can be artificially decreased
to $0$ by blocking all messages. This increases the false
rejection rate though. The same happens with the false rejection
rate: it can be decreased to $0$ by not blocking any messages at all.
A good
spam filter has a low FAR as well as a low FRR. 
We want to be able to rank the performance of spam filters, as expressed by
their accept and reject ratios, using a single scalar value. At first, it would
seem
natural to define the \emph{wrongness} $W$ of a spam filter as the
distance to the origin when plotting the performance of the filter
in the FAR/FRR plane. However we do not consider these errors to be symmetric,
given that it is much worse to falsely reject a legitimate message than
to accept a spam message. We can think of a false positive as an
error, and a false negative as an effectiveness indicator. 
Another way to approach this, is to say that one is willing to tolerate a small
increase in the false reject ratio for a significant reduction of the false
accept ratio. 
Trying
to represent that we propose the function
\begin{equation}
    W(FAR,FRR) = (FRR+\epsilon)^2(FAR+\epsilon)~,
\end{equation}
where $\epsilon$ is a small constant (\ie $\epsilon=.01$).

\subsection{Modelling of the normal email traffic}
\label{sec-modNormTraf}%

The simulator needs to generate
email traffic that faithfully emulates the behaviour of
a normal user, to test the filters with. In order to do that we
need content for the bodies of these emails, and an effective method to
generate normal email traffic patterns.

Ideally, the bodies of the messages should come from real mail
sent by different kinds of users. However, due to privacy
considerations, this is not possible. Instead, we use bodies from
a wide variety of USENET (news) messages. This variety is
necessary to avoid creating email messages with a limited set of
subjects and vocabulary.
We do note however that USENET messages do not contain HTML code (whereas
some of the spam messages considered (see section
\rref{sec-modSpamTraf}) do).

One important issue to model is the fact that the contents of the
messages sent by different people varies. A person's vocabulary
varies as a function of his occupation, education, hobbies, etc.
In order to reflect that in our model, each sender of an email
belongs to a specific topic group chosen randomly. Each topic
group only takes bodies from a specific news group.

Another important issue to model is that in normal email traffic,
people mostly get mail from people they know. To model this
behaviour we enumerate senders and receivers and use a normal
distribution to find people close to the sender of the message.
The use of the normal distribution means that most mail will be
send to people close to the sender in the list, but that there is
still a chance of sending mail to people less close to the sender.

Mailing lists also are considered as a special case, because in
some aspects they behaves like spammers, sending multiple copies
of the same content to many different users, so this behaviour may
confuse some filters. In our model, each mailing list has a email
address database which is initialised at the beginning of the
simulation. When it sends an email, it iterates over its database
sending the same message to all the users in it. When the end of
the database is reached, another message is chosen and the process
starts again.

\subsection{Modelling of the spam traffic}
\label{sec-modSpamTraf}%

To generate realistic spam email-traffic, we have analysed the
behaviour of bulk-mailers.
An important characteristic of spam is that it arrives in bulk.
Whenever a bulk mailer start sending, it keeps sending for some
period of time, until the spam message is delivered to all the
users in it's database.

An issue when generating the spam traffic is to make sure that the
percentage of spam of the total email is realistic. At the end of
2002, 40\%%
\footnote{\url{http://zdnet.com.com/2100-1106-955842.html}}${}^,$%
\footnote{\url{http://www.linuxsecurity.com/articles/privacy_article-6369.html}}
of the total email traffic consist of spam, and is rapidly
growing, accounting for more than 50 percent on June, 2003%
\footnote{\url{http://news.zdnet.co.uk/internet/ecommerce/0,39020372,39118223,00.htm}}.

In our model of spammers, each spammer has a database with email
addresses. When a spammer starts sending, it continues sending as
fast as possible, until the message is delivered to the whole
database. A spammer can send personalised spam. In this case, each
spam message has only one target address, say login@server, and
the line ``Dear login,'' is added to the body of the message.

To generate spam mail we also need content for the bodies of the
spam messages. We use a variety of spam archives to extract these
bodies\footnote{\url{http://www.spamarchive.org/}}. 
In a separate simulation, we also
include a collection of very recent spam messages in order to show
the impact of the spam evolution in the filtering techniques.

\subsection{The simulator}
The simulator must generate authentic-looking, email traffic and
bulk mail patterns. We also believe that the simulator should be
as general as possible. That is, it should be easy to use with
existing filters and with filters that might be invented in the
future. In addition it should also be easy to add new features of
bulk-mailers.

\begin{figure}[!htb]
\begin{center}
\includegraphics[width=8cm]{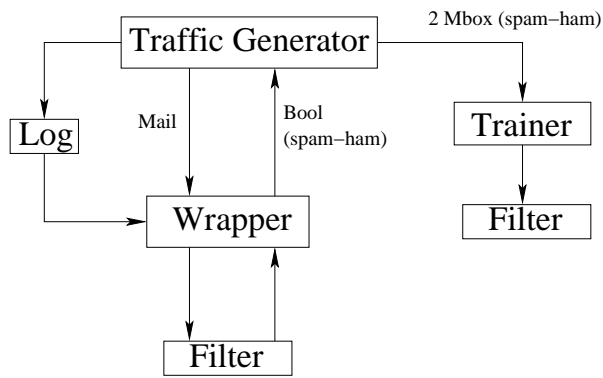}
\caption{The design of the simulator}\label{fig:simulator}
\end{center}
\end{figure}

The architecture chosen to fulfil these requirements is to split
the simulator into four parts (see Fig.~\ref{fig:simulator}).
The first part is the traffic generator. This is the main module,
responsible for generating email traffic according to its
configuration and deliver it. The traffic generator has a set of
senders and receivers. Each sender has a probability to send an
email every one step of the simulation and the traffic generator
iterates over the set of senders in a round-robin fashion. After
the sender is selected, if it is a normal user, the traffic
generator first generates the number of targets this email will
have, in which field (\ie \verb"To:", \verb"Cc:", \verb"Bcc:") and
then selects the receivers. In order to select the receivers, we
first take the index of the sender (in the list of senders)
and then we generate a
offset using the normal distribution. The index plus the offset yields the
receiver. In other words, the smaller
the difference between your index and the index
somebody else, the closer that person is to you, and the more likely it is that
you exchange messages. In the case of a
spammer, it changes to state sending, and it will keep sending
until the spam message is delivered to all the users in it's
database. In case of a mailing list, it behaves like a normal user
except for the fact that their targets are selected from its
database and the same body is sent to the whole database.

The second part is the wrapper, a program that receives an email
on its standard input and is responsible for running the filter,
take its classification (spam or ham) and send the response back
to the traffic generator. If the filter is at the server level,
the traffic generator will log every simulated connection to a log
file, emulating the log file of the mail server. This file is
accessible for the wrapper as well. Each filter may have different
input formats, so the wrapper program is filter-specific.

The third part is the trainer. Given that some filters (\eg
Bayesian) need to be trained before they can be used, and the
email traffic received for each user has different patterns, the
simulator is able to generate a certain amount of correctly
classified traffic, and send it to two different files, one for
ham and one for spam. After that, the trainer is executed, and it
should to be able to train the filter with this data.

\subsection{The analysed filters}
Some of the aforementioned filtering techniques are not included
in our analysis for several reasons. Some filters are highly
dependent on unknown constants (\eg the number of spammers who are
not in a blacklist). By setting those constants to an arbitrary
value, we would basically be setting the results of those
simulations. We also restrict our analysis to open source filters.
We therefore choose to test the following set of popular spam filters.
\begin{description}
\item[Filtering based on traffic analysis]: 
Mail volume-based filter.
\item[Content Based Filters]:
  Distributed Checksum Clearinghouse (DCC),
  Genetic algorithm based spam filter (SpamAssassin), and
  Na\"{\i}ve Bayesian Filters (Bogofilter, Spamprove, Bmf).
\end{description}

\section{Spam filter comparison}
\label{sec-results}

In this section we give for each analysed filter a short
description of the results of the simulation. 
For each of the filters, we have measured their effectiveness both when applied
at user level and at server level, against the following types of spam traffic:
(1) non-personalised spam, (2) personalised spam, and (3) non-personalised
recent spam. 

\begin{table}[t]
\caption{Filter performance: non-personalised spam.}
\label{fig-results}
\begin{center}
\begin{tabular}[b]{lccccc}
\textit{Filter}&\textit{Level}&~\textit{FRR}~ &~\textit{FAR}~ &~$W*10^5$~ \\
\sphline %
Bogofilter  &U     &      .0000 & .144 & 1.56\\%
Bmf         &U     &       .0117 & .029 & 1.86\\%
Mail volume &S     &      .0000 & .633 & 6.44\\%
SpamAssassin&U/S   &         .0071 & .213 & 6.57\\%
DCC         &U/S   &        .0005 & .624 & 7.11\\%
Bmf         &S     &            .0004 & .656 & 7.20\\%
Bogofilter  &S     &       .0000 & .709 & 7.27\\%
Spamprove   &U     & .0085 & .215 & 7.80\\%
Spamprove   &S     &        .0024 & .695 & 10.9\\%
\sphline
\end{tabular}
\end{center}
%
\caption{Filter performance: personalised spam.}
\label{fig-results-per}
\begin{center}
\begin{tabular}[b]{lccccc}
\textit{Filter}&\textit{Level}&~\textit{FRR}~ &~\textit{FAR}~ &~$W*10^5$~ \\
\sphline %
Mail volume &S    &        .0000  & .196 & 2.07 \\%
Bogofilter  &U    &        .0000  & .233 & 2.43 \\%
Bmf         &U    &         .0037  & .186 & 3.70 \\%
Spamprove   &U    &   .0005  & .325 & 3.76 \\%
SpamAssassin&U/S  &           .0070  & .217 & 6.60 \\%
DCC         &U/S  &          .0028  & .710 & 11.9 \\%
\sphline
\end{tabular}
\end{center}
%
%
\caption{Filter performance: non-personalised recent spam.}
\label{fig-results-nspam}
\begin{center}
\begin{tabular}[b]{lccccc}
\textit{Filter}&\textit{Level}&~\textit{FRR}~ &~\textit{FAR}~ &~$W*10^5$~ \\
\sphline %
Bmf           &U      &       .0030 &  .070 & 1.37 \\%
Bogofilter    &U      &      .0000 &  .129 & 1.41 \\%
Spamprove     &U      & .0040 &  .138 & 2.94 \\%
SpamAssassin  &U/S    &         .0074 &  .179 & 5.74 \\%
Mail volume   &S      &      .0000 &  .642 & 6.58 \\%
\sphline
\end{tabular}
\end{center}
\end{table}

The qualitative
results (wrongness, FAR and FRR) are summarised in several graphs and tables.
The tables contain the performance of the filters both at user and at server
level, indicated in the \emph{Level} column by U and S respectively. 
If the performance of the filter did not depend on its level, U/S is used to
indicate this fact.
Note that in case a Bayesian filter runs at server level, it did not get a
per-user training but a general one.
Table~\ref{fig-results} contains the results for non-personalised spam. 
Table~\ref{fig-results-per} shows the results of the simulation
when personalised spam was used. Note that in this case the filters were
only tested when running at user level.
Finally, for Table~\ref{fig-results-nspam}
non-personalised spam was used but this time the spam mails were
collected during the last two months. The purpose of this is to
show the impact of the new spamming techniques (\eg like including
random words in the body of the message), especially on the
Bayesian filters. The performance of the filters is also shown graphically in
Fig.~\ref{fig-graph1}--\ref{fig-graph3}.

In the remainder of this section we discuss the performance and behaviour of
each of the tested filters individually, in separate subsections.

\newcommand{\cput}[3]{\put(#1,#2){\vbox to 0pt{\vss\hbox to 0pt{\hss#3\hss}\vss}}}

\begin{figure}[t]
\setlength{\unitlength}{4cm}
\sidebyside{
\begin{picture}(1.4,1.4)(-0.2,0)
\put(0,0){\line(1,0){1.1}} %
\put(0.94,-0.19){\textit{FRR}}%
\put(-0.07,-0.07){$0$}%
\put(0.47,-0.08){$.01$}%
\put(0.98,-0.08){$.02$}%
\put(0.1,0){\line(0,1){0.01}}%
\put(0.2,0){\line(0,1){0.01}}
\put(0.3,0){\line(0,1){0.01}}
\put(0.4,0){\line(0,1){0.01}}
\put(0.5,0){\line(0,1){0.02}}
\put(0.6,0){\line(0,1){0.01}}
\put(0.7,0){\line(0,1){0.01}}
\put(0.8,0){\line(0,1){0.01}}
\put(0.9,0){\line(0,1){0.01}}
\put(1,0){\line(0,1){0.03}}
\put(0,0){\line(0,1){1.1}}
\put(-0.15,0.92){\rotatebox{90}{\textit{FAR}}} %
\put(-0.08,0.48){$.5$}%
\put(-0.07,0.98){$1$} \put(0,0.1){\line(1,0){0.01}}
\put(0,0.2){\line(1,0){0.01}} \put(0,0.3){\line(1,0){0.01}}
\put(0,0.4){\line(1,0){0.01}} \put(0,0.5){\line(1,0){0.02}}
\put(0,0.6){\line(1,0){0.01}} \put(0,0.7){\line(1,0){0.01}}
\put(0,0.8){\line(1,0){0.01}} \put(0,0.9){\line(1,0){0.01}}
\put(0,1){\line(1,0){0.03}}
\cput{0}{0.63396}{$\bullet$}%
\cput{0.029}{0.62491}{$\star$}%
\cput{0.358}{0.21303}{$*$}%
\cput{0.0015}{0.14458}{$\otimes$}%
\cput{0.429}{0.21587}{$\circledcirc$}%
\cput{0.586}{0.02947}{$\oplus$}%
\cput{0.0025}{0.70976}{$\times$}%
\cput{0.1215}{0.69501}{$\circ$}%
\cput{0.02}{0.65614}{$+$}%
\end{picture}
\vspace{2mm}
\caption{Non-personalised spam.}
\label{fig-graph1}
}
{
%
%
\setlength{\unitlength}{4cm}
\begin{picture}(1.4,1.4)(-0.2,0)
\put(0,0){\line(1,0){1.1}} %
\put(0.94,-0.19){\textit{FRR}}%
\put(-0.07,-0.07){$0$}%
\put(0.47,-0.08){$.01$}%
\put(0.98,-0.08){$.02$}%
\put(0.1,0){\line(0,1){0.01}}%
\put(0.2,0){\line(0,1){0.01}} \put(0.3,0){\line(0,1){0.01}}
\put(0.4,0){\line(0,1){0.01}} \put(0.5,0){\line(0,1){0.02}}
\put(0.6,0){\line(0,1){0.01}} \put(0.7,0){\line(0,1){0.01}}
\put(0.8,0){\line(0,1){0.01}} \put(0.9,0){\line(0,1){0.01}}
\put(1,0){\line(0,1){0.03}}
\put(0,0){\line(0,1){1.1}}
\put(-0.15,0.92){\rotatebox{90}{\textit{FAR}}} %
\put(-0.08,0.48){$.5$}%
\put(-0.07,0.98){$1$} \put(0,0.1){\line(1,0){0.01}}
\put(0,0.2){\line(1,0){0.01}} \put(0,0.3){\line(1,0){0.01}}
\put(0,0.4){\line(1,0){0.01}} \put(0,0.5){\line(1,0){0.02}}
\put(0,0.6){\line(1,0){0.01}} \put(0,0.7){\line(1,0){0.01}}
\put(0,0.8){\line(1,0){0.01}} \put(0,0.9){\line(1,0){0.01}}
\put(0,1){\line(1,0){0.03}}
\cput{0.000}    {0.19698}{$\bullet$}%
\cput{0.142}    {0.71054}{$\star$}%
\cput{0.352}    {0.21716}{$\ast$}%
\cput{0.0000}   {0.23301}{$\otimes$}%
\cput{0.0295}   {0.32534}{$\circledcirc$}%
\cput{0.185}    {0.18642}{$\oplus$}%
\end{picture}
\vspace{2mm}
\caption{Personalised spam.}
\label{fig-graph2}
}
%
%
\sidebyside{
\setlength{\unitlength}{4cm}
\begin{picture}(1.4,1.4)(-0.2,0)
\put(0,0){\line(1,0){1.1}} %
\put(0.94,-0.19){\textit{FRR}}%
\put(-0.07,-0.07){$0$}%
\put(0.47,-0.08){$.01$}%
\put(0.98,-0.08){$.02$}%
\put(0.1,0){\line(0,1){0.01}}%
\put(0.2,0){\line(0,1){0.01}} \put(0.3,0){\line(0,1){0.01}}
\put(0.4,0){\line(0,1){0.01}} \put(0.5,0){\line(0,1){0.02}}
\put(0.6,0){\line(0,1){0.01}} \put(0.7,0){\line(0,1){0.01}}
\put(0.8,0){\line(0,1){0.01}} \put(0.9,0){\line(0,1){0.01}}
\put(1,0){\line(0,1){0.03}}
\put(0,0){\line(0,1){1.1}}
\put(-0.15,0.92){\rotatebox{90}{\textit{FAR}}} %
\put(-0.08,0.48){$.5$}%
\put(-0.07,0.98){$1$} \put(0,0.1){\line(1,0){0.01}}
\put(0,0.2){\line(1,0){0.01}} \put(0,0.3){\line(1,0){0.01}}
\put(0,0.4){\line(1,0){0.01}} \put(0,0.5){\line(1,0){0.02}}
\put(0,0.6){\line(1,0){0.01}} \put(0,0.7){\line(1,0){0.01}}
\put(0,0.8){\line(1,0){0.01}} \put(0,0.9){\line(1,0){0.01}}
\put(0,1){\line(1,0){0.03}}
\cput{0.002}    {0.64251}   {$\bullet$} 
\cput{0.371}    {0.17924}   {$\ast$} 
\cput{0.0035}   {0.12933}   {$\otimes$} 
\cput{0.0204}   {0.13815}   {$\circledcirc$} 
\cput{0.154}    {0.07006}   {$\oplus$} 
\end{picture}
\vspace{2mm}
\caption{Non-personalised recent spam.}
\label{fig-graph3}
}
{\begin{tabular}[b]{lcc}\\
\textit{Filter}&\textit{Level}&~\textit{Symbol}~ \\
\sphline %
Bogofilter  &S     &$\times$ \\
Bogofilter  &U     &$\otimes$ \\
Bmf         &S     &$+$ \\
Bmf         &U     &$\oplus$ \\
Spamprove   &S     &$\circ$ \\
Spamprove   &U     &$\circledcirc$ \\
Mail volume &S     &$\bullet$ \\
SpamAssassin&U/S   &$\ast$ \\
DCC         &U/S   &$\star$ \\
\sphline
\end{tabular}
}
\end{figure}
\subsection{Mail volume-based filter}
The volume-based filter uses an algorithm that checks how much
email is received from a specific host during the last
connections (the last 1500 lines from the log file in the
simulation, which is the same as used by Kai's
Spamshield\footnote{\url{http://spamshield.conti.nu/}}). If the
amount of mail received is greater then a certain threshold, then
the mail is classified as spam. This filter was able to correctly
classify all the legitimate emails, for a high enough threshold.
The drawback of this filter is that the FAR achieved is in general
very high.

When personalisation is used, this filter performs very well. 
We consider this to be
a side effect, because for personalised mail 
the bulk mailer has to establish many
more connections in order to deliver the same amount of emails.
That makes it easier to detect with mail volume analysis. 
 Detection is easily avoided however using multiple
open-relays at the same time.

\subsection{Distributed Checksum Clearinghouse}
The Distributed Checksum Clearinghouse filter tested is a
modification of the standard DCC filter
version~1.2.14\footnote{\url{http://www.rhyolite.com/anti-spam/dcc/}}.
The modification disables the report function to the Internet
server. Instead, a local database was used and the filter reports
the emails to this database.

The percentage of false positives of the DCC filter is small, but
the performance filtering spam is small as well. This performance
can be improved by a less conservative threshold, but this has a
direct impact on the FRR.

When personalised spam is used the accuracy of this filter lowers
slightly.

\subsection{Genetic algorithm based spam filter}
The genetic trained filter that we used in our simulations was
SpamAssassin
version~2.60\footnote{\url{http://spamassassin.org/}}. The default
filter configuration was used in our simulation.

This filter performs very well, achieving one of the best
performances of the evaluated filters at ISP level, with the
non-personalised spam. This accuracy is not largely affected when
personalised spam is used.
One drawback of this filter is that it is the most
computationally expensive of the evaluated filters.

\subsection{Na\"{\i}ve Bayesian Filters}
Several implementations of Bayesian filters were evaluated and
there were important performance differences between them. An
outstanding performance was achieved by Bogofilter when it runes
at user level, it had almost no false positives and more than 75\%
of the spam was filtered in every simulation. We suspect that this
good performance is due to the Fisher's method, see
Robinson~\cite{robinson}. Bmf has also a very good performance,
comparable to Bogofilter, but less conservative. It archives
lowest FAR of the evaluated filters, but the number of false
positives in some circumstances is certainly high. Spamprove has a
low efficacy compared with other Bayesian filters. We suspect this
is related to the fact that Spamprove ignores HTML code.

A general characteristic of Bayesian filters is that their
performance is lowered drastically when they run at ISP level.
Another general characteristic of Bayesian filters is that most of
the wrongly classified emails are very short. We suspect that in
these cases there is not enough information to perform
statistical analysis.

Personalisation of the spam does not have a big impact on
the accuracy of Bayesian filters. If anything, it improves the
accuracy of some of them. We suspect this improvement is related
to the fact that keywords like ``login'' in the body of the
message becomes a good spam indicator.

\section{Conclusions}

\paragraph{Filtering at the ISP level} The most efficient way for filtering  at the ISP
level seems to be using a genetic algorithm. This require a big
amount of processing power, but when that is available it is
certain a good option. A mail volume-based filter can be
established as a first line of defence even when its performance
is low because for a high enough threshold they give no false
rejects and are computationally cheap.

\paragraph{Filtering at the user level} The best way for a user to
filter spam seems to be a na\"{i}ve Bayesian filter with the
Fisher's method like Bogofilter. Depending on how important it is
for the user to loose email, Bmf also can be considered as a good
option.

The simulator is in the public domain, and can be downloaded from
\url{http://www.cs.kun.nl/~flaviog/spam-filter/}.


\chapbblname{spam-filter} \chapbibliography{spam-filter}

\end{document}